\documentclass[%
 reprint,
superscriptaddress,
 amsmath,amssymb,
 aps,
pra,
]{revtex4-2}

\usepackage{graphicx}
\usepackage{dcolumn}
\usepackage{bm}
\usepackage{natbib}
\usepackage{textcase}
\usepackage{url}
\usepackage{amsmath}
\usepackage{framed}

\begin{document}

\preprint{APS/123-QED}

\title{Extrusion of chromatin loops by a composite loop extrusion factor}

\author{Hao Yan}%
\affiliation{%
	Integrated Graduate Program in Physical and Engineering Biology, Yale University, New Haven, Connecticut 06520, USA
}
\affiliation{%
 Department of Physics, Yale University, New Haven, Connecticut 06520, USA
}%
\author{Ivan Surovtsev}%
\affiliation{%
 Department of Physics, Yale University, New Haven, Connecticut 06520, USA
}%
\affiliation{%
	Department of Cell Biology, Yale School of Medicine, New Haven, Connecticut 06520, USA
}%
\author{Jessica F Williams}%
\affiliation{%
	Department of Cell Biology, Yale School of Medicine, New Haven, Connecticut 06520, USA
}%
\author{Mary Lou P Bailey}
\affiliation{%
	Integrated Graduate Program in Physical and Engineering Biology, Yale University, New Haven, Connecticut 06520, USA
}
\affiliation{%
	Department of Applied Physics, Yale University, New Haven, Connecticut 06520, USA
}%

\author{Megan C King}%
\affiliation{%
	Integrated Graduate Program in Physical and Engineering Biology, Yale University, New Haven, Connecticut 06520, USA
}
\affiliation{%
	Department of Cell Biology, Yale School of Medicine, New Haven, Connecticut 06520, USA
}%
\affiliation{%
	Department of Molecular, Cellular, and Developmental Biology, Yale University, New Haven, Connecticut 06520, USA
}%
\author{Simon G J Mochrie}%
 \email{simon.mochrie@yale.edu}
 \affiliation{%
	Integrated Graduate Program in Physical and Engineering Biology, Yale University, New Haven, Connecticut 06520, USA
}
\affiliation{%
 Department of Physics, Yale University, New Haven, Connecticut 06520, USA
}%
\affiliation{%
	Department of Applied Physics, Yale University, New Haven, Connecticut 06520, USA
}%


%

\date{\today}
             
\begin{abstract}             
Chromatin loop extrusion by Structural Maintenance of Chromosome (SMC) complexes is thought to underlie
intermediate-scale chromatin organization inside cells.
Motivated by a number of experiments suggesting that nucleosomes may block loop extrusion by SMCs,
such as cohesin and condensin complexes,
we introduce and characterize theoretically a composite loop extrusion factor (composite LEF) model.
In addition to an SMC complex that creates a chromatin loop by encircling two threads of DNA,
this model includes
a remodeling complex that relocates or removes nucleosomes as it progresses along the chromatin, and
nucleosomes that block SMC translocation along the DNA.
Loop extrusion is enabled by
 SMC motion along nucleosome-free  DNA,  created in the wake of the remodeling complex,
 while
nucleosome re-binding behind the SMC acts as a ratchet, holding the SMC close to the remodeling complex.
We show that, for a wide range of parameter values, this collection of factors constitutes a composite LEF that extrudes loops with a velocity, comparable to the velocity of remodeling complex translocation on chromatin in the absence of SMC, and much faster than loop extrusion by an isolated SMC that is blocked by nucleosomes.    
\end{abstract}

\maketitle

\section{\label{sec:level1}Introduction}

Exquisite spatial organization is a defining property of chromatin, allowing the genome both to be accommodated within the volume of the cell nucleus, and  simultaneously accessible to the transcriptional machinery, necessary for gene expression. On the molecular scale, histone proteins organize 147~bp of DNA into nucleosomes, that are separated one from the next by an additional 5-60~bp \cite{Linker}.
On mesoscopic scales ($10^5-10^6$~bp),
it has long been understood that  loops are an essential feature of chromatin organization.
The recent development of chromosome conformation capture (Hi-C) techniques now enables quantification of chromatin organization via a proximity ligation assay, that yields a map of the relative probability that any two genomic locations are in contact with each other \cite{Dekker2002}.
Hi-C contact maps
have led to the identification of topologically associating domains (TADs) as fundamental elements of intermediate-scale chromatin organization \cite{RN22,RN144,RN16,RN2,RN143}.
Genomic regions inside a TAD interact frequently with each other, but have relatively little contact with regions in even neighboring TADs.

Although, how TADs arise remains uncertain, the loop extrusion factor (LEF) model
has emerged as the preferred candidate mechanism for 
TAD formation.
In
this model, LEFs -- identified as the Structural Maintenance of Chromosome (SMC) complexes,  cohesin and condensin --
encircle two chromatin threads,
forming the base of a loop, and then initiate loop extrusion
\cite{Alipour:2012aa,RN11,Fudenberg:2016aa,RN153,Goloborodko:2016aa,Goloborodko:2016ab}.
Efficient topological cohesin loading onto chromatin,
as envisioned by the LEF model,
depends both on the presence of the Scc2-Scc4 cohesin loading complex
and on cohesin's {ATP}-ase activity \cite{Murayama2013}.
Loop extrusion proceeds until the LEF is blocked by another LEF or until it encounters a boundary element,
generally identified
as  DNA-bound CCCTC-binding factor (CTCF),
or until it dissociates, causing the corresponding loop to dissipate.
Thus, a population of LEFs leads to a dynamic steady-state chromatin organization.
As may be expected, based on the correlation between TAD boundaries and CTCF binding sites
\cite{Khoury2020},
this model recapitulates important features of
experimental Hi-C contact maps  \cite{Alipour:2012aa,Fudenberg:2016aa,RN153}.

The LEF model was recently
bolstered by beautiful single-molecule experiments that directly visualized DNA loop extrusion
by condensin \cite{RN147} and cohesin  \cite{RN169}. 
However, both of these studies focused on the behavior of the SMC complex on naked DNA,
whereas inside cells DNA is densely decorated with nucleosomes.
Ref.~\cite{RN169} (and  then Ref.~\cite{Kong2020})
did also show that cohesin could compact lambda DNA (48,000 bp) loaded with about three nucleosomes, but this nucleosome density  ($6 \times 10^{-5}$~bp$^{-1}$)  is nearly
100-fold less than the nucleosome density in chromatin ($5 \times 10^{-3}$~bp$^{-1}$).

\begin{figure*}
\includegraphics[width=0.89\textwidth]{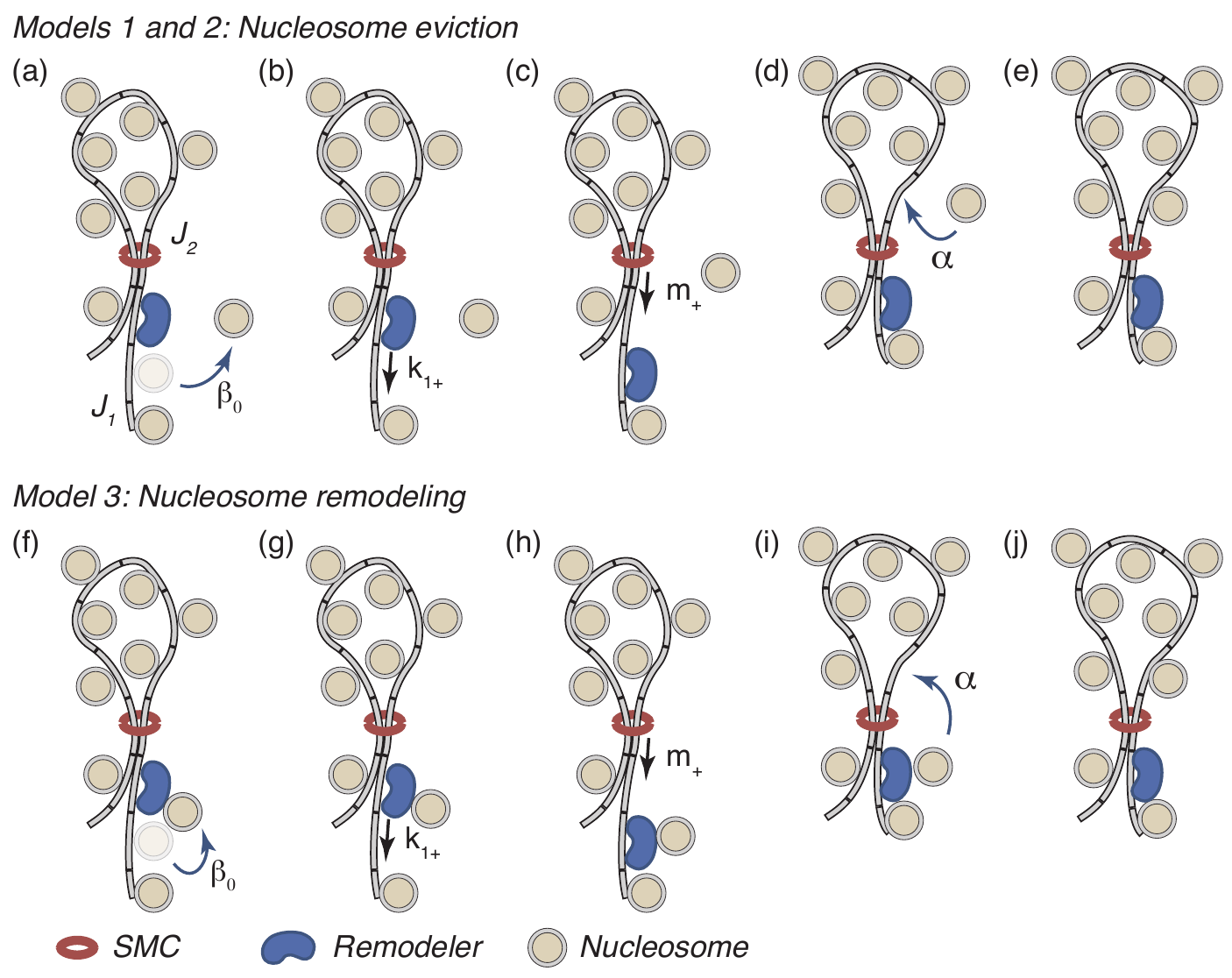}
\caption{
Loop extrusion via a composite LEF, comprising an SMC complex, which forms a ring around two nucleosome-free sections of DNA, nucleosomes that block SMC translocation, and a remodeling complex which removes nucleosomes in front of the SMC.
In our model, a single loop extrusion step starts when the remodeling complex
forces a nucleosome from the DNA ahead of the remodeler,
thus moving the junction (J1) between nucleosomal DNA and naked DNA one step forward. $\beta_0$ is the rate of
nucleosome dissociation (a) or remodeling (f) when the remodeler is next to a nucleosome.
Next, the remodeler moves into the resultant nucleosome-free region,  (b) and (g).
$k_{1+}$ is the rate at which the remodeler steps forward, when the remodeler-nucleosome separation is one
step.
Then, the SMC complex moves into the new nucleosome-free region left behind the remodeler
 (c) and (h).
 $m_+$ is the rate at which the the SMC steps forward on nucleosome-free DNA.
Finally, a nucleosome rebinds behind the SMC complex,
moving the second junction (J2) between nucleosomal DNA and naked DNA one step forward, and
so preventing the SMC from subsequently backtracking.
$\alpha$ is the rate of nucleosome rebinding (d) or re-formation (i).
After these four sub-steps,
the LEF configuration is the same as before the first step, but
the loop is one step larger, (e) and (j).
The top row (a-e) illustrates a hypothetical scenario (models 1 and 2) in which the displaced nucleosome is in solution
before rebinding DNA behind the SMC.
The bottom row (f-j) illustrates an alternative ``remodeled-nucleosome"
scenario (model 3) in which the displaced nucleosome
remains associated with the remodeling complex 
before rebinding DNA behind the SMC.
}
\label{LEFcartoon}
\end{figure*}

The notion that nucleosomes might actually represent a barrier for SMC translocation and therefore loop extrusion
is suggested by
measurements that reveal that cohesin motions on nucleosomal DNA are much reduced compared to those on naked DNA \cite{RN159}.
Further supporting the hypothesis that nucleosomes hinder SMC-driven
loop extrusion are several studies indicating that cohesin translocation requires transcription-coupled nucleosome remodeling \cite{RN145,RN180,RN182,RN183,RN181}.
In particular, Ref. \cite{RN181}
demonstrates that cohesin,
recruited to one genomic location by a cohesin loading complex,
is relocated to another
by RNA polymerase (Pol II) during transcription.
Finally, Ref.~ \cite{Golfier821306} found that
presence of nucleosomes in  {\em Xenopus laevis} egg extract prevented DNA exposed to the extract from looping and compaction.

In this paper,
motivated by the possibility that nucleosomes block loop extrusion by SMCs,
we introduce and characterize theoretically a composite loop extrusion factor (composite LEF) model
that realizes chromatin
loop extrusion.
Fig.~\ref{LEFcartoon} is a cartoon representation of
this model.
As illustrated in the figure, in addition to
an SMC complex that
encircles two threads of DNA,
creating a chromatin loop, the model includes
a remodeling complex,
that removes or relocates nucleosomes as it translocates along chromatin,
and nucleosomes, that create a barrier for
SMC motion. 
Both the remodeler and the nucleosomes are
 essential components of the composite LEF.
 
We envision that when the ring-like SMC complex is threaded by DNA, it can move along the DNA
until it encounters a nucleosome,
which blocks its motion.
We hypothesize that
the SMC's ATPase activity does not exert enough force to move a nucleosome,
even though it may give rise to directional loop extrusion on naked DNA.
Without nucleosome remodeling, therefore,
an SMC complex remains trapped  by its surrounding nucleosomes at a more-or-less fixed genomic location.
Directional loop extrusion is enabled by
SMC motion along the nucleosome-free thread of DNA, that is created in the wake of the remodeling complex, and is maintained by the SMC being held close to the remodeling complex by the ratcheting action of nucleosomes re-locating
to behind the SMC.
The composite LEF, illustrated in Fig.~\ref{LEFcartoon},
extrudes the right-hand thread of the chromatin loop,
embraced by its constituent SMC complex.
The left-hand thread of the loop remains encircled by the SMC at a fixed genomic location,
with the SMC trapped by its surrounding nucleosomes.
In our model,
two-sided loop extrusion would require a remodeler on each thread.
The model is agnostic concerning the specific identity of the remodeler,
except that it must be able to displace nucleosomes
or alter their configuration in a manner that allows the SMC to subsequently pass them by.
The top row of Fig.~\ref{LEFcartoon} illustrates a hypothetical process,
in which the displaced nucleosome unbinds from ahead of the remodeler,
before the same or a different nucleosome subsequently rebinds behind the SMC.
The bottom row illustrates an alternative version of the model,
in which the displaced nucleosome remains associated with the LEF in a transient, ``remodeled'' configuration,
that is permissive to loop extrusion.

 This paper is organized as follows.
 In Sec.~\ref{Theory}, we calculate the velocities of one-sided loop extrusion for three,
 slightly different versions of the composite LEF model.
 In fact, differences among the loop extrusion velocities of the different models are small.
 In Sec.~\ref{Discussion}, we examine the results of Sec.~\ref{Theory}  to elucidate the conditions required for
 efficient loop extrusion. We also  compare the composite LEF's loop extrusion velocity
 to the velocities of the remodeler and  the SMC, each translocating alone on chromatin.
For a broad range of parameter values,
we find that the model's component factors can indeed be sensibly identified as a composite LEF,
that can extrude chromatin loops at a velocity that is comparable to
 that of isolated remodeler translocation on chromatin,
 and much faster than loop extrusion by an isolated SMC, that is blocked
 by nucleosomes.
 Finally, in Sec.~\ref{Conclusion}, we conclude.

\section{Theory}
\label{Theory}
The results presented in this section rely on and were guided by
the calculations and ideas presented in
Refs.~\cite{Peskin1993,Betterton:2003aa,PhysRevE.71.011904},
concerning other examples of biological Brownian ratchets.
To calculate the loop extrusion velocity, $v$,
in terms of the rates of remodeling complex forward ($k_+$)  and backward ($k_-$) stepping on DNA,
the rates of SMC forward ($m_+$) and backward ($m_-$) stepping on DNA, and the rates of nucleosome binding ($\alpha$) and unbinding ($\beta$), {\em etc.},
we make a number of simplifying assumptions.
First, we consider chromatin as a sequence of nucleosome binding sites.
Second, we assume that the none of the SMC complex, the remodeling complex,
and nucleosomes can occupy the same location, {\em i.e.} we assume an infinite hard-core
repulsion between these factors, that prevents their overlap.
Third, we assume that 
there are 
 well-defined junctions between bare DNA and nucleosomal DNA
 in front of the remodeler (junction 1) and
behind the SMC loop  (junction 2), so that
when a remodeler forces a nucleosome from junction 1,  subsequently it relocates to junction 2.
Finally, we hypothesize that, although SMCs can not push nucleosomes out of their way,
the remodeling complex can.
Following  Refs.~\cite{Betterton:2003aa,PhysRevE.71.011904},
we actualize this nucleosome-ejecting activity via a nearest-neighbor repulsive interaction,
$\Delta G$, between the remodeling complex and junction 1.

\subsection{Model 1}
First, we consider a streamlined model (model 1),
which assumes that
the nucleosome unbinding and re-binding rates
are much faster than the remodeling complex and SMC forward-
and backward-stepping rates.
Because of this separation of time scales,
we can consider that the SMC and remodeler move in a free energy landscape
defined by the time-averaged configuration of nucleosomes  \cite{Peskin1993}.
Thus, when the remodeling complex and junction 1 are next to each other (zero separation),
the free energy is $\Delta G$, corresponding to the nearest-neighbor remodeler-junction repulsive interaction,
or,
when there are $n$ nucleosome binding sites between the remodeling complex
and junction 1, the free energy is $n \Delta g$, corresponding to the free energy of
$n$ unbound nucleosomes in front of the remodeling complex.
A straightforward equilibrium statistical mechanical calculation then informs us that
the probability that the remodeling complex and junction 1 are not next to each other
is
\begin{equation}
P_1 =
 \frac{1}
 {
 1 + e^\frac{\Delta g-\Delta G}{k_B T}-e^{-\frac{\Delta G}{k_B T}}
 }.
 \label{EQP7P7}
 \end{equation}
Similarly, the probability that the SMC and junction 2 are not next to each other is
  \begin{equation}
  P_2=e^{-\frac{\Delta g}{k_B T}},
  \label{EQP3P3}
  \end{equation}
because we assume there is not a SMC-nucleosome nearest-neighbor interaction,
beyond the requirement that they not be at the same location.

The principle of detailed balance informs us that the ratio of 
forward and backward transition
rates are given by a Boltzmann factor.
Therefore,
when the
the remodeling complex and junction 1 are not next to each other, we expect
\begin{equation}
\frac{k_{+}}{k_{-}} =  e^{-\frac{\Delta G_R}{k_BT}},
\label{BackAndForth2-}
\end{equation}
where $\Delta G_R$ is the free energy change involved in moving the remodeler one step forward.
However, when
the remodeling complex and junction 1 are next to each other, this ratio of rates
is modified, because of the nucleosome-remodeling complex repulsion:
\begin{equation}
\frac{k_{1+}}{k_{0-}} =  e^{-\frac{\Delta G}{k_BT}} \frac{k_+}{k_-},
\label{BackAndForth2}
\end{equation}
where 
$k_{1+}$ is the remodeling complex forward stepping rate, when the remodeler-junction 1 separation is one step,   
and
$k_{0-}$ is the remodeling complex backward stepping rate when the remodeler-junction 1 separation is zero.
As discussed in detail in
Refs.~\cite{Betterton:2003aa,PhysRevE.71.011904}, to satisfy  Equation~(\ref{BackAndForth2}),
in general, we can write
\begin{equation}
k_{1+}= e^{-\frac{\Delta G f}{k_BT}} k_+,
\label{EQ1EQ1}
\end{equation}
\begin{equation}
k_{0-} =  e^{\frac{\Delta G (1-f)}{k_BT}} k_-,
\label{EQ2EQ2}
\end{equation}
where  $0<f<1$ \cite{Betterton:2003aa,PhysRevE.71.011904}.
However, as discussed in detail in
Refs.~ \cite{Betterton:2003aa} and \cite{PhysRevE.71.011904} in an analogous context, the choice $f=0$
maximizes the composite LEF velocity.
Therefore, we pick $f=0$, so that
\begin{equation}
k_{1+}=k_+
\label{RatesNo1}
\end{equation}
and
\begin{equation}
k_{0-} =  e^{\frac{\Delta G}{k_BT}} k_-,
\label{RatesNo2}
\end{equation}
which satisfy Equation~(\ref{BackAndForth2}).
Then,
the mean velocity of the remodeling complex
may be written
\begin{eqnarray}
  v_R 
  & = b k_+ P_1 -b k_- P_3 P_1-b k_- e^\frac{\Delta G}{k_BT} P_3 (1-P_1)
\label{F18-VVVold}
 \end{eqnarray}
 where
 $P_3$ is the probability that the remodeling complex and the SMC are not next to each other, and
$b$ is the step size along the DNA, taken to be the separation between nucleosomes for simplicity.
The first term on the right-hand side of Equation~(\ref{F18-VVVold})
corresponds to stepping forward, which can only happen if the remodeling complex and junction 1 are not next to each other.
The second term on the right-hand side of Equation~(\ref{F18-VVVold})
corresponds to stepping backwards in the case that the remodeling complex and junction 1 are not next to each other and the remodeling complex and the
SMC complex are not next to each other, in which case the
rate of this process is $k_-$.
The third term on the right-hand side of Equation~(\ref{F18-VVVold})
corresponds to stepping backwards in the case that the remodeling complex and junction 1 are next to each other and the remodeling complex and the
SMC complex are not next to each other, in which case the
rate of this process is $k_- e^\frac{\Delta G}{k_B T}$,
according to Equation~(\ref{RatesNo2}).
Using Equation~(\ref{EQP7P7})
in Equation~(\ref{F18-VVVold}), we find
 \begin{eqnarray}
  v_R 
&= b \frac{
 {k_+}-k_- e^\frac{\Delta g}{k_B T} P_3}
{
1 + e^\frac{\Delta g-\Delta G}{k_B T}-e^{-\frac{\Delta G}{k_B T}}
}.
\label{F18-VVV}
 \end{eqnarray}
We can also calculate the diffusivity of the remodeler:
 \begin{eqnarray}
 D_R &= \frac{1}{2} \left ( b^2 k_+ P_1 + b^2  k_- P_3 P_1 + b^2 k_- e^\frac{\Delta G}{k_BT} P_3 (1-P_1)     \right ) \nonumber \\
 &= \frac{1}{2} b^2 \frac{
 {k_+}  + k_- e^\frac{\Delta g}{k_B T} P_3}
{
1 + e^\frac{\Delta g-\Delta G}{k_B T}-e^{-\frac{\Delta G}{k_B T}}
}.
 \end{eqnarray}
Similar reasoning informs us that the velocity and diffusivity of the SMC complex  are
  \begin{equation}
 v_S  = b (m_ + P_3   -m_-  P_2)
 = b (m_ + P_3   -m_-  e^{-\frac{\Delta g}{k_B T}}),
\label{F18-VVVV}
 \end{equation}
 and
   \begin{equation}
 D_S
 = \frac{1}{2} b^2 (m_ + P_3    + m_-  e^{-\frac{\Delta g}{k_B T}}),
\label{F18-DDDD}
 \end{equation}
 respectively.

Equation~(\ref{F18-VVV}) shows that the velocity of the remodeling complex, $v_R$, decreases with increasing $P_3$,
while Equation~(\ref{F18-VVVV}) shows that the velocity of the SMC complex, $v_S$, increases with increasing $P_3$.
To realize a composite LEF, $P_3$ must take on a value that causes these two velocities to coincide, so that
the remodeling complex and the SMC complex translocate together
with a common velocity, $v$, given by $v=v_R=v_S$.
Equations~(\ref{F18-VVV}) and  (\ref{F18-VVVV}) constitute two equations for the
two unknowns,  $P_3$ and $v $. Solving  yields
\begin{eqnarray}
P_3
 & =\frac{ \frac{ k_+}{k_- m_+} +(e^{-\frac{\Delta g}{k_BT}}-e^{-\frac{\Delta g+\Delta G}{k_BT}} +e^{-\frac{\Delta G}{k_BT}})\frac{m_- }{k_- m_+}}
{
\frac{ e^\frac{\Delta g}{k_BT} }{m_+}
 +
\frac{1+e^\frac{\Delta g-\Delta G}{k_BT} -e^{-\frac{\Delta G}{k_BT} }
 }{k_-}
 }  
\label{EQ28}
\end{eqnarray}
and
\begin{eqnarray}
 v
&=b
 \frac{ 
 \frac{k_+}{k_-}-\frac{m_-}{m_+}}
{
\frac{
e^\frac{\Delta g}{k_BT}}{m_+}
 +\frac{1+e^\frac{\Delta g-\Delta G}{k_BT} -e^{-\frac{\Delta G}{k_BT}} }{k_-}}.
 \label{EQ29}
\end{eqnarray}
Using this value for $P_3$,
it further follows that
\begin{eqnarray}
D_R =& b^2
\frac{
k_+
}{
1+e^\frac{\Delta g-\Delta G}{k_B T}-e^{-\frac{\Delta G}{k_BT}}
}
\nonumber \\
&+
\frac{1}{2}b^2
\frac
{
\frac{m_-}{m_+} -\frac{k_+}{k_-}
}
{
	\frac
	{
	e^\frac{\Delta g}{k_B T}
	}
	{
	m_+
	} 
+
	\frac
	{
	1+e^\frac{\Delta g-\Delta G}{k_B T} -e^{-\frac{\Delta G}{k_B T}}
	}
	{
	k_-
	}
}
\label{EQ-S23}
\end{eqnarray}
and
\begin{eqnarray}
D_S
= &\frac{1}{2} b^2 
\frac{ \frac{ k_+}{k_- } +(e^{-\frac{\Delta g}{k_BT}}-e^{-\frac{\Delta g+\Delta G}{k_BT}} +e^{-\frac{\Delta G}{k_BT}})\frac{m_- }{k_-}}
{
\frac{ e^\frac{\Delta g}{k_BT} }{m_+}
 +
\frac{1+e^\frac{\Delta g-\Delta G}{k_BT} -e^{-\frac{\Delta G}{k_BT} }
 }{k_-}
 }  
 \nonumber \\ &+ \frac{1}{2} b^2 e^{-\frac{\Delta g}{k_B T}} m_-.
 \label{EQ-S24}
\end{eqnarray}

\begin{figure}
\begin{center}
\includegraphics[width=0.45\textwidth]{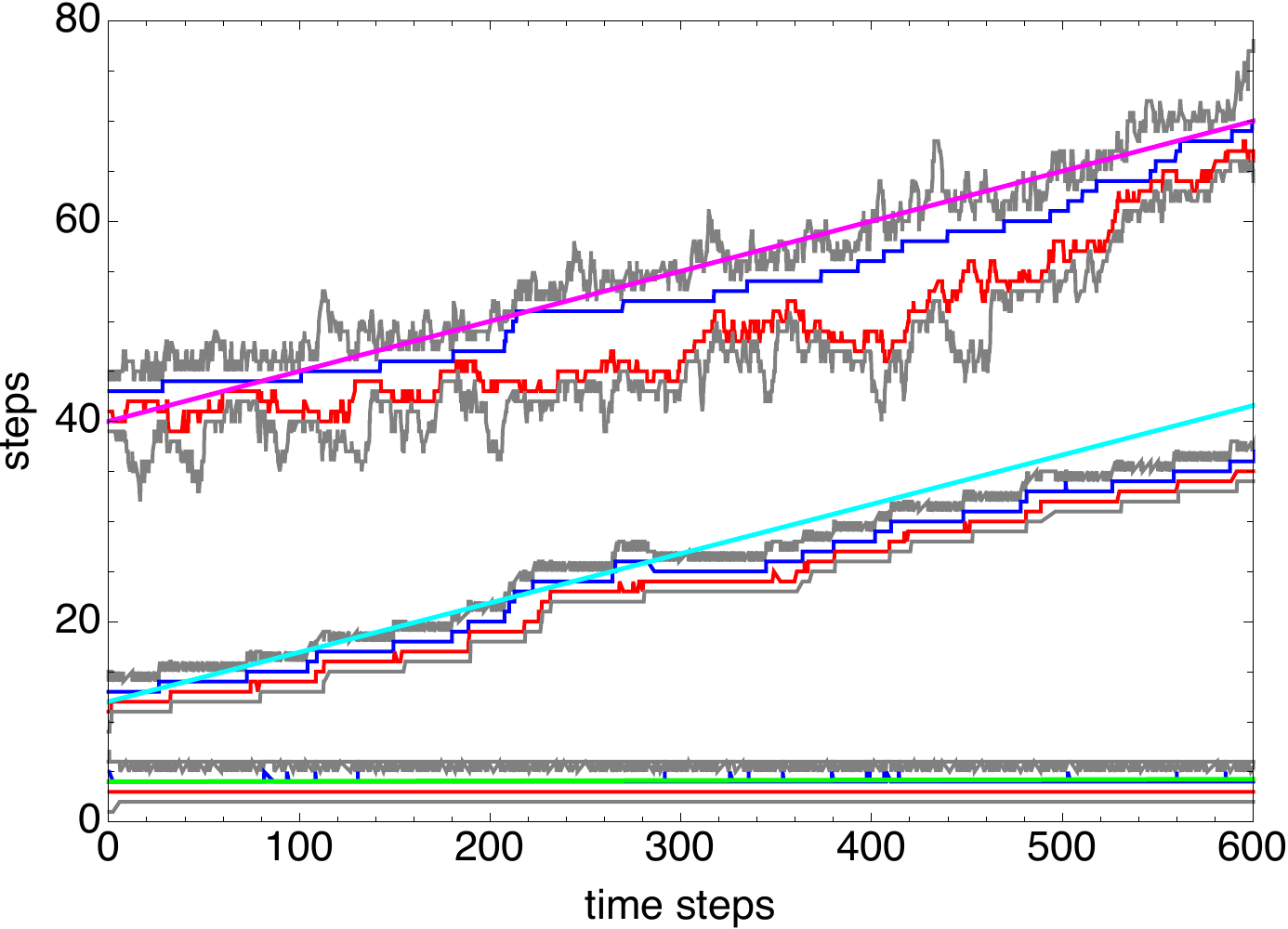}
\end{center}
\caption{
Three example composite LEF trajectories from model 2 simulations.
In each case, the positions versus time of the nucleosome junctions are shown gray, the remodeling complex  is shown blue,
and the SMC complex is shown red.
When tracking together, each such group of four traces constitutes a composite LEF.
The model parameters are
$k_+ = 0.05$ per time step,
$k_- = 5 \times 10^{-7}$ per time step,
$m_+ = m_- = 0.3$ per time step,
$\Delta G = 18.0 k_BT$,
$\alpha = 1$ per time step, and
$\beta - \alpha e^\frac{-\Delta g}{k_B T}$
for all three composite LEFs, but
$\Delta g = 18.0 k_B T$  for the bottom
group of traces, 
$\Delta g = 9.0 k_B T$ for the middle group of traces,
and  $\Delta g = 0.5 k_B T$ for the top group of traces.
The cyan, green, and magenta lines each have a slope given by the theoretical composite LEF velocity
for the parameters of each simulation.
}
\label{Figure2}
\end{figure}
 
\subsection{Model 2}
At the cost of a little complication, it is possible to calculate the composite LEF velocity, 
even when the nucleosome binding ($\alpha$) and unbinding ($\beta$) rates are not
much larger than $k_+$, $k_-$, $m_+$, and $m_-$.
This model  (model 2) is preferable {\em a priori}
 because we expect the nucleosome unbinding rate, $\beta$, to be small.
 In fact, the results obtained with model 2 are very similar to those obtained with model 1.

Similar to the remodeling complex forward- and backward-stepping rates, when the remodeling complex and junction 1
are adjacent,
the nucleosome binding and unbinding rates are modified as follows:
\begin{equation}
\frac{\alpha_{1}}{\beta_{0}} =  e^{-\frac{\Delta G}{k_BT}} \frac{\alpha}{\beta},
\label{BackAndForth1}
\end{equation}
where $\alpha_1$ is the nucleosome binding rate when the remodeler-junction 1 separation
is one step, and $\beta_0$ is the nucleosome unbinding rate when the remodeling complex and junction 1 are adjacent (separation 0).
To satisfy Equation~(\ref{BackAndForth1}),
we can write
\begin{equation}
\alpha_{1}= e^{-\frac{\Delta G f}{k_BT}} \alpha,
\end{equation}
and
\begin{equation}
\beta_{0} =  e^{\frac{\Delta G(1-f)}{k_BT}} \beta,
\end{equation}
which stand
alongside
Equations~(\ref{RatesNo1}) and (\ref{RatesNo2}).
As above,
we again choose $f=0$, so that
\begin{equation}
\alpha_{1}=\alpha
\end{equation}
and
\begin{equation}
\beta_{0} =  e^{\frac{\Delta G}{k_BT}} \beta.
\end{equation}
To proceed in this case, we first write down the
mean velocity of junction 1:
\begin{eqnarray}
  v_{J1}
&=  b ( \beta-\alpha)P_1 +  b \beta e^{\frac{\Delta G}{k_B T}} (1-P_1),
\label{LEF-P1}
 \end{eqnarray}
where
$P_1$ is the probability that the remodeling complex and junction 1 are not next to each other and
 $b$ is the step size.
Similarly, we can also write down the mean velocity of the remodeling complex:
 \begin{equation}
  v_R  = 
 b (k_+-k_- P_3 )P_1
 -b   k_- e^{\frac{\Delta G}{k_B T}} (1-P_1)  P_3,
\label{LEF-P2}
 \end{equation}
 where $P_3$ is the probability that the remodeling complex and the SMC complex are
 not adjacent to each other. 
 Next, we write down the velocity of the SMC complex:
 \begin{equation}
  v_S  = 
b ( m_+ P_3 -m_- P_2), 
\label{LEF-P3}
 \end{equation}
 where  $P_2$ is the probability that the SMC complex and the junction between
 bare DNA and nucleosomal DNA  behind the SMC complex, namely junction 2,
 are not adjacent to each other.
 Finally, we can write down the mean velocity of junction 2:
  \begin{equation}
  v_{J2}  = 
b ( \alpha P_2 -\beta).
\label{LEF-P4}
 \end{equation}

\begin{figure*}
\begin{center}
\includegraphics[width=0.78\textwidth]{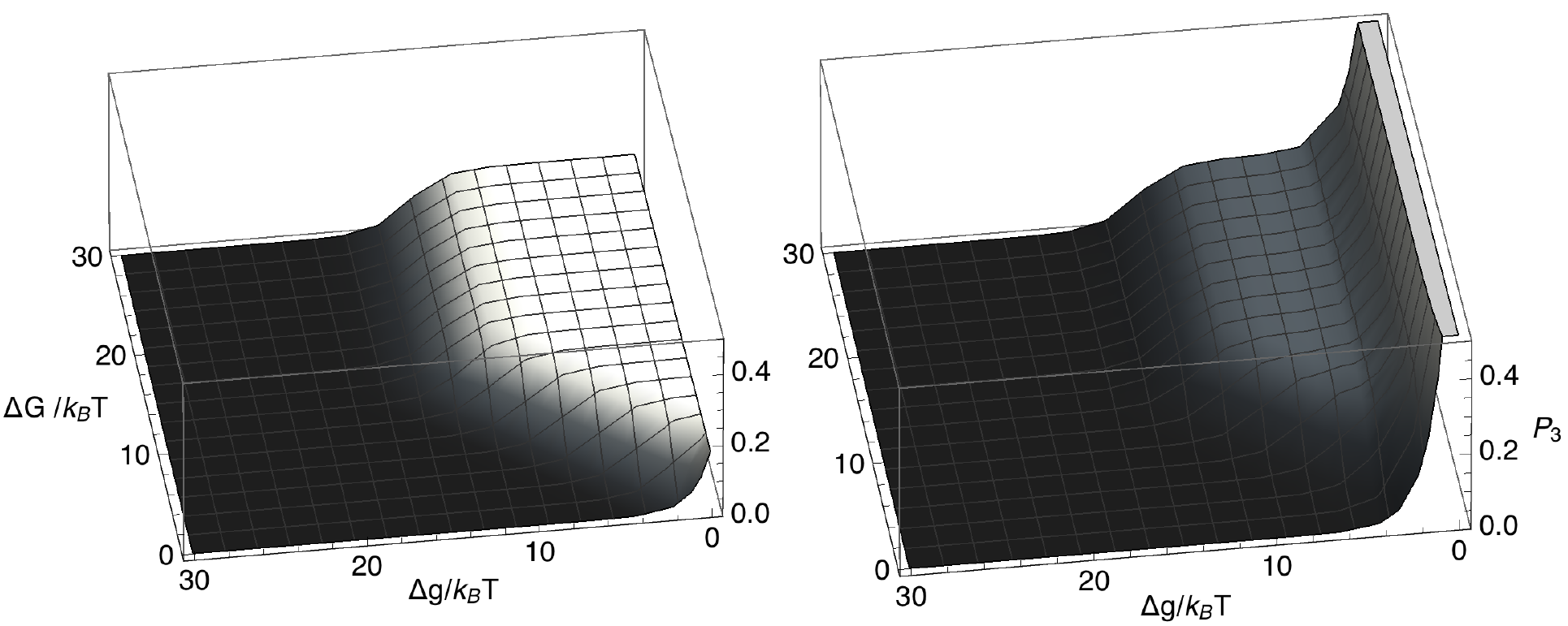}
\end{center}
\caption{
Probability, $P_3$, that the remodeling complex and the SMC complex involved in a composite LEF are not adjacent to each other, plotted versus $\Delta G/(k_B T)$ and $\Delta g/(k_BT)$, according to model 1
[Equation (\ref{EQ28})] for $k_+ = 0.05$~ per time step, $k_- = 5.0 \times 10^{-7}$ per time step,
$m_+ = 0.3$~per time step, and $m_- = 0.0003$~per time step (left) or
$m_- = 0.3$ per time step (right).
}
\label{Figure3}
\end{figure*}
 
For the composite LEF to translocate as a single entity,
 it is necessary for each of its component parts to translocate with a common velocity, $v$, where
 \begin{equation}
v=  v_{J1}  =   v_R  =   v_S  =  v_{J2}.
 \label{LEF-Vs}
 \end{equation}
 Solving Equations~(\ref{LEF-P1}) through (\ref{LEF-Vs}) for the four unknowns, namely $v$, and the probabilities, $P_1$, $P_3$, and $P_2$,
 yields the values of these quantities.
To this end, first we solve
Equations~(\ref{LEF-P1}) and (\ref{LEF-P2}), assuming that junction 1 and the remodeling complex have
a common velocity ($v_1$) with the result that
\begin{equation}
 v_1  = 
b \frac{
( \frac{k_+}{k_- }-\frac{\alpha}{\beta} P_3)
}
{
(1-e^{-\frac{\Delta G}{k_BT}})(\frac{1}{k_- }   + \frac{P_3}{\beta} )+\frac{\alpha +k _+}{\beta k_-} e^{-\frac{\Delta G}{k_BT}}
}.
\label{EQ-LEF16}
\end{equation}
Next, we solve
Equations~(\ref{LEF-P3}) and (\ref{LEF-P4}), assuming that the SMC complex and junction 2
have a common velocity ($v_2$). In this case, we find
\begin{equation}
 v_{2}  =
b
\frac{
\frac{m_+ P_3}{m_-} -\frac{\beta}{\alpha}
}
{
\frac{1}{m_-}+\frac{1}{\alpha}
}
\label{EQ-LEF17}
\end{equation}
for the SMC-junction 2 velocity.
For these two pairs to translocate together, manifesting a four component, composite LEF,
it is necessary that they share a common velocity, $v$, given by $v = v_1  =  v_2$.
Setting Equation~(\ref{EQ-LEF16}) equal to Equation~(\ref{EQ-LEF17}) and solving for $P_3$, we find
\begin{widetext}
\begin{align}
P_3 =&
\frac{1}{2 m_+ (1-e^{-\frac{\Delta G}{k_B T}})}
\left ( -\alpha-m_- 
-\frac{m_+}{k_-} (\alpha + k_+)e^{-\frac{\Delta G}{k_B T}}
 + ( \frac{\beta m_-}{\alpha}- \frac{\beta m_+}{k_+} ) (1-e^{-\frac{\Delta G}{k_B T}}) \right )
\nonumber \\
&+  \frac{\beta m_-}{2 m_+ (1-e^{-\frac{\Delta G}{k_B T}})}
\sqrt{
\begin{aligned}
    \frac{ 4 m_+}{ \beta m_-}
    (1-e^{-\frac{\Delta G}{k_B T}}) ( \frac{k_+}{\alpha k_-} + \frac{k_+}{k_- m_-} + 
      \frac{ \beta}{\alpha k_-} (1-e^{-\frac{\Delta G}{k_B T}})
      +(\frac{k_+}{\alpha k_-}+ \frac{1}{ k_-}) e^{-\frac{\Delta G}{k_B T}}) \\
     + \left ( \frac{1}{\beta} + \frac{\alpha}{\beta m_-} 
     +(\frac{\alpha m_+}{\beta k_- m_-} + \frac{k_+ m_+}{\beta k_- m_-} ) e^{-\frac{\Delta G}{k_B T}}
      +
     (\frac{m_+}{k_- m_-}-\frac{1}{\alpha}) (1-e^{-\frac{\Delta G}{k_B T}}) \right )^2.
    \end{aligned}
  }
  \label{EQ-finalP2}
    \end{align}
           \end{widetext}
 The velocity of the composite LEF can be calculated by substituting
Equation~(\ref{EQ-finalP2}) into
Equation~(\ref{EQ-LEF17}).

We also carried out a series
of Gillespie simulations  \cite{Gill77} of model 2 for several values of $\frac{\Delta g}{k_B T}$.
Fig.~\ref{Figure2} shows the position versus time for
three example simulations,  each carried out for a different value of $\frac{\Delta g}{k_B T}$.
For each of these LEFs, the gray traces represent the positions of the junctions between nucleosomal DNA and naked DNA,
the blue trace represents the position of the remodeling complex, and
the red trace represents the position of the SMC complex.
The mean position of the bottom LEF, which corresponds to $\frac{\Delta g}{k_B T}=18$,
remains essentially fixed over the period of the simulation, implying a very small LEF velocity.
In addition, in this case,
the remodeler and the SMC remain next to each other throughout the trajectory,
implying a very small value of $P_3$.
By contrast, the mean position of the middle
LEF  ($\frac{\Delta g}{k_B T}=8$) 
 increases more-or-less linearly in time with the remodeler and the SMC both stepping forward and
frequently moving out of contact. Thus, in this case,  the LEF shows a significant velocity
and an intermediate value of $P_3$.
Finally, although the velocity of the top LEF  ($\frac{\Delta g}{k_B T}=0.5$)
is very similar  to that of the middle LEF, 
the top LEF shows many fewer
remodeler-SMC contacts than the middle LEF, corresponding to a
significantly larger value of $P_3$.
The cyan, green, and magenta lines in Fig.~\ref{Figure2} 
 have slopes given by the corresponding model-2 composite LEF velocities, 
-- calculated by substituting
Equation~(\ref{EQ-finalP2}) into
Equation~(\ref{EQ-LEF17}) --
revealing good agreement between theory and simulation.

\subsection{Model 3}
Model 3
 supposes that the probability of complete nucleosome unbinding into solution is negligible,
but that there exists a ''remodeled'' configuration, in which the nucleosome is both associated with the remodeler and also sufficiently displaced to  allow the
remodeler to step forward (bottom row of Fig.~\ref{LEFcartoon}).
In this model, we interpret $\Delta g$ to be the free energy of the remodeled configuration.
For simplicity,  we also assume
a separation of time scales with remodeling occurring much faster than translocation.
Then, the probability that the
remodeling complex and junction 1 are not next to each other is
\begin{equation}
P_1=
\frac{1}
{
1+e^\frac{\Delta g -\Delta G}{k_B T}
}
=
\frac{1}
{
1+(1+e^\frac{\Delta g}{k_B T}) e^\frac{\Delta G}{k_B T} - e^\frac{\Delta G}{k_B T}
},
\label{EQ32}
\end{equation}
while
the probability that the SMC and junction 2
are not next to each other is
\begin{equation}
P_2=
\frac{1}
{
1+e^\frac{\Delta g}{k_B T}
}.
\label{EQ33}
\end{equation}
Equations~(\ref{EQ32}) and (\ref{EQ33})
replace model 1's
Equations~(\ref{EQP7P7}) and (\ref{EQP3P3}), respectively.
However, Equations~(\ref{F18-VVVold}) and (\ref{F18-VVVV}) are unchanged  for model 3.
It is apparent therefore that we may write down the model-3 results for $P_3$ and $v$
by replacing $e^\frac{\Delta g}{k_B T}$ in corresponding results for model 1 by
 $e^\frac{\Delta g}{k_B T} +1$.
Thus, for model 3, we find
 \begin{eqnarray}
P_3
 & =\frac{ \frac{ k_+}{k_- m_+} +( 
 \frac{
 1-e^{-\frac{\Delta G}{k_BT}}
 }
 {
 1+ e^{\frac{\Delta g}{k_BT}}
 } + e^{-\frac{\Delta G}{k_BT}})\frac{m_- }{k_- m_+}}
{
\frac{1+ e^\frac{\Delta g}{k_BT}  }{m_+}
 +
\frac{1+e^\frac{\Delta g-\Delta G}{k_BT}
 }{k_-}
 }  
\label{EQ28-model3}
\end{eqnarray}
and
\begin{eqnarray}
 v
&=b
 \frac{ 
 \frac{k_+}{k_-}-\frac{m_-}{m_+}}
{
\frac{
1+ e^\frac{\Delta g}{k_BT}}{m_+}
 +\frac{1+e^\frac{\Delta g-\Delta G}{k_BT}
 }{k_-}}.
 \label{EQ29-model3}
\end{eqnarray}

\section{Discussion}
\label{Discussion}
To realize a composite LEF, junction 1 and the remodeler,
 on the one hand, must not outrun the SMC and junction 2, on the other.
This requirement may be expressed mathematically by insisting that
the probability, $P_3$, that the remodeling complex
and the SMC are not next to each other, must be less than 1.
Otherwise, for $P_3=1$,  the remodeler and SMC do not come into contact, and
 we may infer that the remodeler has outpaced the SMC.
Fig.~\ref{Figure3}  plots
 $P_3$,
according to model 1,
as a function of $\frac{\Delta G}{k_B T}$ and $\frac{\Delta g}{k_B T}$.
For the parameter values, used in the left-hand panel, we see that $P_3$ 
is everywhere less than 1, consistent with the existence of a composite LEF throughout the region illustrated.
In fact, $P_3$
takes on a relatively large plateau value for
\begin{equation}
\Delta G > \Delta g
\label{EQ15}  
\end{equation}
 and
 \begin{equation}
m_+ > e^\frac{\Delta g}{k_BT} k_-.
\label{EQ16}
\end{equation}
Elsewhere, $P_3$ is small.

For the parameter values used in the right-hand panel of Fig.~\ref{Figure3}, however,
although $P_3$ shows a similar plateau at intermediate values of
$\frac{\Delta g}{k_B T}$,
as $\frac{\Delta g}{k_B T}$ decreases to near zero,
 $P_3$ increases rapidly to unity, and 
according to Equation~(\ref{EQ28}),
would unphysically exceed unity for small enough $\frac{\Delta g}{k_B T}$.
This circumstance arises when even $P_3=1$ is not sufficient to satisfy $v_R = v_S$.
When the remodeling complex and junction 1 outrun the SMC and junction 2 -- {\em i.e.}  
when $v_R > v_S$ --
the premise of a composite LEF, upon which Equations~(\ref{EQ28}) and (\ref{EQ29}) are based, can no longer hold.
Thus, to achieve a composite LEF, we must have that $v_R \leq v_S$ for $P_3 = 1$.
This condition requires that the model parameter values must satisfy
\begin{equation}
m_ +    -m_-  e^{-\frac{\Delta g}{k_B T}} >
\frac{
 {k_+}-k_- e^\frac{\Delta g}{k_B T}}
{
1 + e^\frac{\Delta g-\Delta G}{k_B T}-e^{-\frac{\Delta G}{k_B T}}
}.
\label{CompositeLEFCondition}
\end{equation}
This condition is violated at small $\Delta g$ for the parameters used in the right-hand panel of
Fig.~\ref{Figure3}.
For $k_+ \gg k_- e^{\frac{\Delta g}{k_B T}}$ and $m_+ \gg m_- e^{-\frac{\Delta g}{k_B T}}$,
the condition for a composite LEF to exist
becomes simply $m_+ > k_+$, namely the forward stepping rate of the SMC on naked DNA should be larger than the
forward stepping rate of the remodeler on naked DNA.

\begin{figure}[t]
\begin{center}
{\includegraphics[width=0.42\textwidth,keepaspectratio=true]{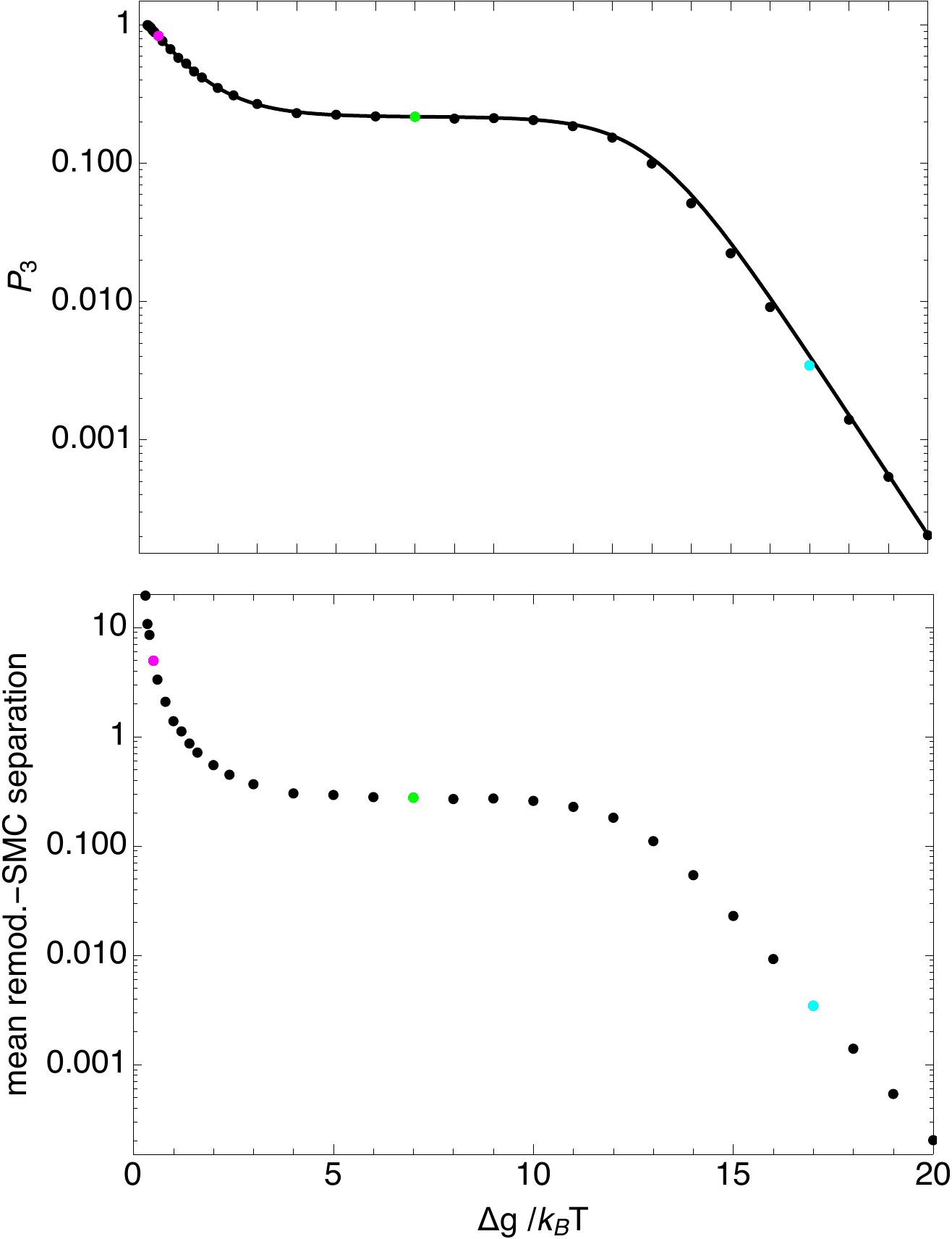}}
\end{center}
\caption{
The probability, $P_3$,  that the remodeler and SMC are not next to each other (top)
and the mean remodeler-SMC separation (bottom),  plotted versus nucleosome binding energy,
$\frac{\Delta g}{k_B T}$. The circles correspond to results  determined from model-2 Gillepsie simulations, each
containing $2^{20}$ transitions.
The solid line corresponds to Equation~(\ref{EQ-finalP2}).
The parameter values used were:
$k_+ = 0.05$ per time step,
$k_- = 5 \times 10^{-7}$ per time step,
$m_+ = m_- = 0.3$ per time step,
$\Delta G = 18.0 k_BT$,
$\alpha = 1$ per time step, and
$\beta = \alpha e^{-\frac{\Delta g}{k_B T}}$.
These parameters correspond to those for Fig.~\ref{Figure2}.
The cyan, green, and magenta points at $\frac{\Delta g}{k_B T} = 0.5$, $8.0$, and $18$, respectively, correspond to the bottom, middle, and
top traces of Fig.~\ref{Figure2}.
\label{Figure4}
}
\end{figure}

To further elucidate the composite LEF's behavior as $P_3$ increases, we turned to Gillespie simulations of the
sort illustrated in Fig.~\ref{Figure2}.
The points in
Fig.~\ref{Figure4} show the simulated results for both $P_3$ itself (top panel)
and the remodeler-SMC separation (bottom panel), plotted
versus $\frac{\Delta g}{k_B T}$.
The solid line in the top panel corresponds to Equation~(\ref{EQ-finalP2}),
demonstrating excellent quantitative agreement between
theory and simulation for $P_3$.
For the parameters of Fig.~\ref{Figure4},  as $\frac{\Delta g}{k_B T}$ decreases below about 3,
$P_3$  increases from its plateau value,
eventually reaching unity at  $\frac{\Delta g}{k_B T} \simeq 0.2$.
Thus, in this case, for $\frac{\Delta g}{k_B T} < 0.2$, a composite LEF does not exist.

It is apparent from the bottom panel of Fig.~\ref{Figure4}, that
the remodeler-SMC separation matches $P_3$ for $\frac{\Delta g}{k_B T} \geq 3$.
This result obtains because, for $\frac{\Delta g}{k_B T} \geq 3$,
the overwhelmingly prevalent remodeler-SMC separations are 0 and 1, so that the calculation of
$P_3$ and the calculation of the mean remodeler-SMC separation are effectively the same calculation in this regime.
However, as $\frac{\Delta g}{k_B T}$ decreases below 3, the mean remodeler-SMC separation rapidly increases
beyond $P_3$, as larger remodeler-SMC separations than 1 become prevalent, as may seen for  the top LEF in Fig.~\ref{Figure2}, which corresponds to $\frac{\Delta g}{k_B T} = 0.5$.
The mean remodeler-SMC separation reaches 1 for $\frac{\Delta g}{k_B T} \lesssim 1.3$ and rapidly increases as
$\frac{\Delta g}{k_B T}$ decreases further.

A key assumption of our theory is that displaced nucleosomes rebind only at
junctions between nucleosomal DNA and naked DNA. However, when the model predicts a relatively large region of naked DNA between the remodeler
and the SMC, into which a nucleosome could easily fit, this assumption seems likely to be inappropriate and the model no longer self-consistent, in turn
suggesting that the condition specified by Equation~(\ref{CompositeLEFCondition}) may be too permissive.
However, further investigation of this question lies beyond the simple model described here.

\begin{figure*}
\begin{center}
\includegraphics[width=0.78\textwidth]{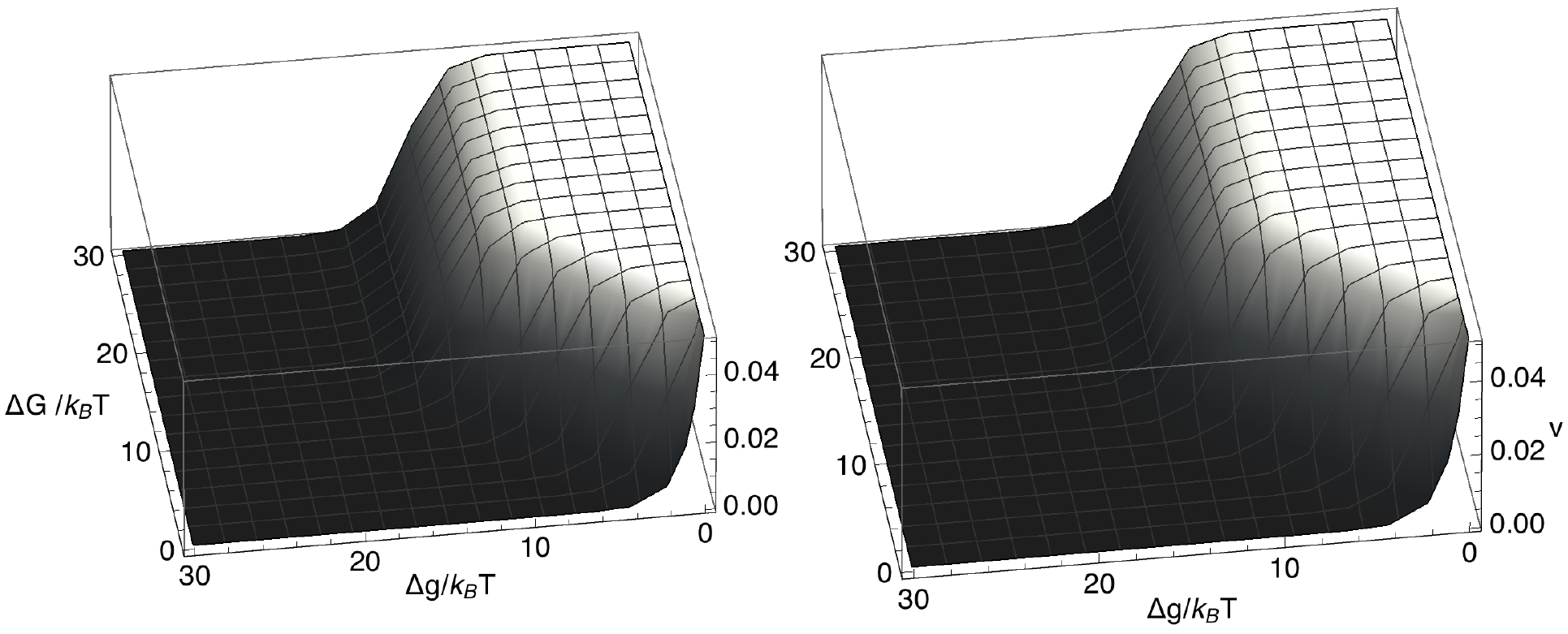}
\end{center}
\caption{
Mean velocity, $v$, of a composite LEF plotted versus $\frac{\Delta G}{k_B T}$ and $\frac{\Delta g}{k_B T}$, according
to Equation~(\ref{EQ29}) for $k_+ = 0.05$ per time step, $k_- = 5.0 \times 10^{-7}$ per time step,
$m_+ = 0.3$ per time step, and $m_- = 0.0003$ per time step (left) or $m_- = 0.3$ per time step (right).
}
\label{Figure5}
\end{figure*}

Fig.~\ref{Figure5} plots the model-1 LEF velocity,
corresponding to the probabilities displayed in Fig.~\ref{Figure3},
showing that
$v$ achieves a relatively large plateau value when the conditions,
\begin{equation}
\Delta G > \Delta g
\label{EQ15}  
\end{equation}
 and
 \begin{equation}
m_+ > e^\frac{\Delta g}{k_BT} k_-,
\label{EQ16}
\end{equation}
 are both satisfied.
 Equation~(\ref{EQ15}) informs us that to achieve rapid composite LEF translocation,
 a large repulsive nucleosome-remodeling complex interaction ($\Delta G$) is necessary, that
 overcomes the nucleosome binding free energy ($\Delta g$).
We might have expected that rapid composite LEF translocation would also require that
 the rate at which the SMC complex steps forward into a gap between the SMC complex and the remodeling complex must exceed the rate at which the remodeling complex steps backwards into that same gap, which is $e^\frac{\Delta G}{k_BT}k_- $, {\em i.e.} we might have expected that $m_+ > e^\frac{\Delta G}{k_BT} k_-$.
However,   because of Equation~(\ref{EQ15}),
Equation~(\ref{EQ16}) is actually
a weaker condition on $m_+$  than this expectation.

\begin{figure*}
\begin{center}
\includegraphics[width=0.78\textwidth]{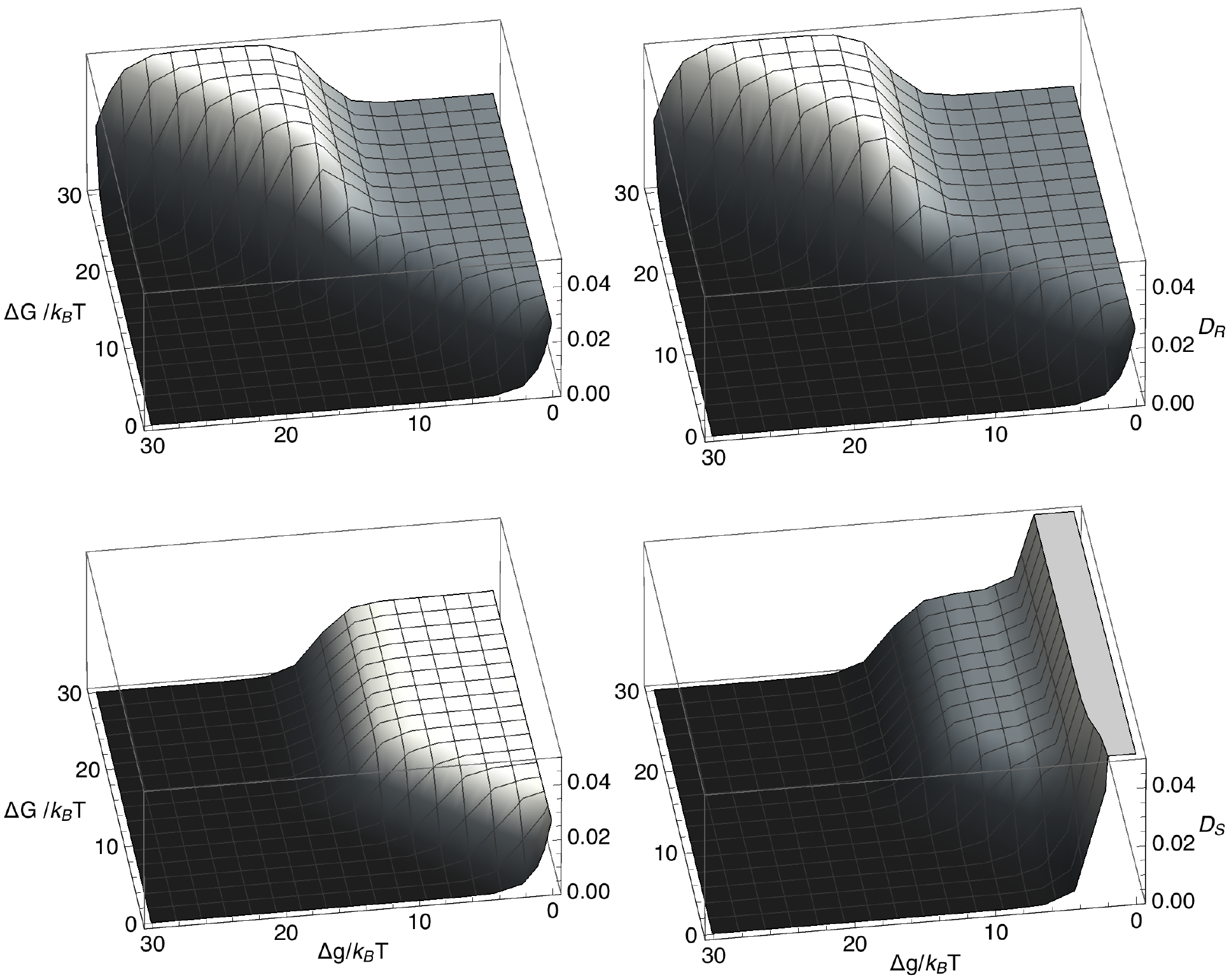}
\end{center}
\caption{
Diffusivities, $D_R$ (top row) and $D_S$ (bottom row) of the remodeling complex and
the SMC complex, respectively, plotted versus $\Delta G/(k_BT)$ and $\Delta g/(k_BT)$,
according to
model 1 [Equations~(\ref{EQ-S23}) and (\ref{EQ-S24})] for $k_+ = 0.05$ per time step, $k_- = 5.0 \times 10^{-7}$ per time step,
$m_+ = 0.3$ per time step, and $m_- = 0.0003$ per time step (left column) or $m_- = 0.3$ per time step (right column).
}
\label{Figure6}
\end{figure*}

Fig.~\ref{Figure6} illustrates the model 1 diffusivities of the remodeling complex and the SMC complex.
Each diffusivity specifies the corresponding factor's positional fluctuations, about the mean displacement, determined by the velocity.
The diffusivities also show relatively large plateau values when Equations~(\ref{EQ15}) and (\ref{EQ16}) are satisfied.
Surprisingly, the diffusivity of the remodeling complex also shows a second plateau
with an even higher plateau value for $\Delta G > \Delta g$ and
$m_+ < e^\frac{\Delta g}{k_BT} k_-$, where the corresponding composite LEF velocity is small.

When all of Equations~(\ref{CompositeLEFCondition}), (\ref{EQ15}) and (\ref{EQ16}) are
simultaneously satisfied,
the plateau values of the 
the probability that the remodeling complex and the SMC complex are not next to each other,
the LEF velocity, and the two diffusivities are given approximately by
\begin{equation}
P_3 \simeq \frac{k_+}{m_+},
\label{plateau1}
\end{equation}
\begin{equation}
v \simeq
b k_-
( \frac{k_+}{k_-} -\frac{m_-}{m_+}   ),
\label{plateau2}
\end{equation}
\begin{equation}
D_R \simeq \frac{1}{2} b^2 k_- (\frac{k_+}{k_-}+\frac{ m_-}{m_+} ),
\end{equation}
and
\begin{equation}
D_S \simeq \frac{1}{2} b^2 k_+,
\end{equation}
respectively.
The plateau value of the composite LEF's loop extrusion velocity is
independent of $\Delta g$.
This result is possible (although not required -- see below) because a loop extrusion step does not lead to a net change in
the nucleosome configuration.

Fig.~\ref{Figure5} shows that the LEF velocity is inevitably small for small $\Delta G$.
For $\Delta G=0$,  corresponding to
solely hard-core repulsions between the remodeler and a nucleosome --
what could be termed a ``passive'' composite LEF,
in analogy to the passive helicase, discussed for example in Ref.~\cite{PhysRevE.71.011904} --
Equation~(\ref{EQ29}) becomes
\begin{equation}
v = b e^{-\frac{\Delta g}{k_BT}  }
 \frac{ 
 \frac{k_+}{k_-} -\frac{m_-}{m_+}}
{ \frac{1}{m_+}  +\frac{1 }{k_-}}.
 \label{EQ30}
\end{equation}
In this case, the composite LEF velocity decreases exponentially with the free energy of nucleosome unbinding,
$\Delta g$.
Since $\Delta g$ is several tens of $k_BT$, we do not expect this limit to be feasible
for effective loop extrusion.
Although Equation~(\ref{EQ30}) corresponds to $f=0$ and $\Delta G =0$, it may be shown that it
also gives the LEF velocity for $f=1$ in the large-$\Delta G$ limit.
This is because for $f = 1$,  large $\Delta G$ effectively creates a hard wall for the remodeler,  albeit located one step away from the nucleosome,  recapitulating the situation considered for $f=0$ and $\Delta G = 0$.

\begin{figure*}
\begin{center}
\includegraphics[width=0.78\textwidth]{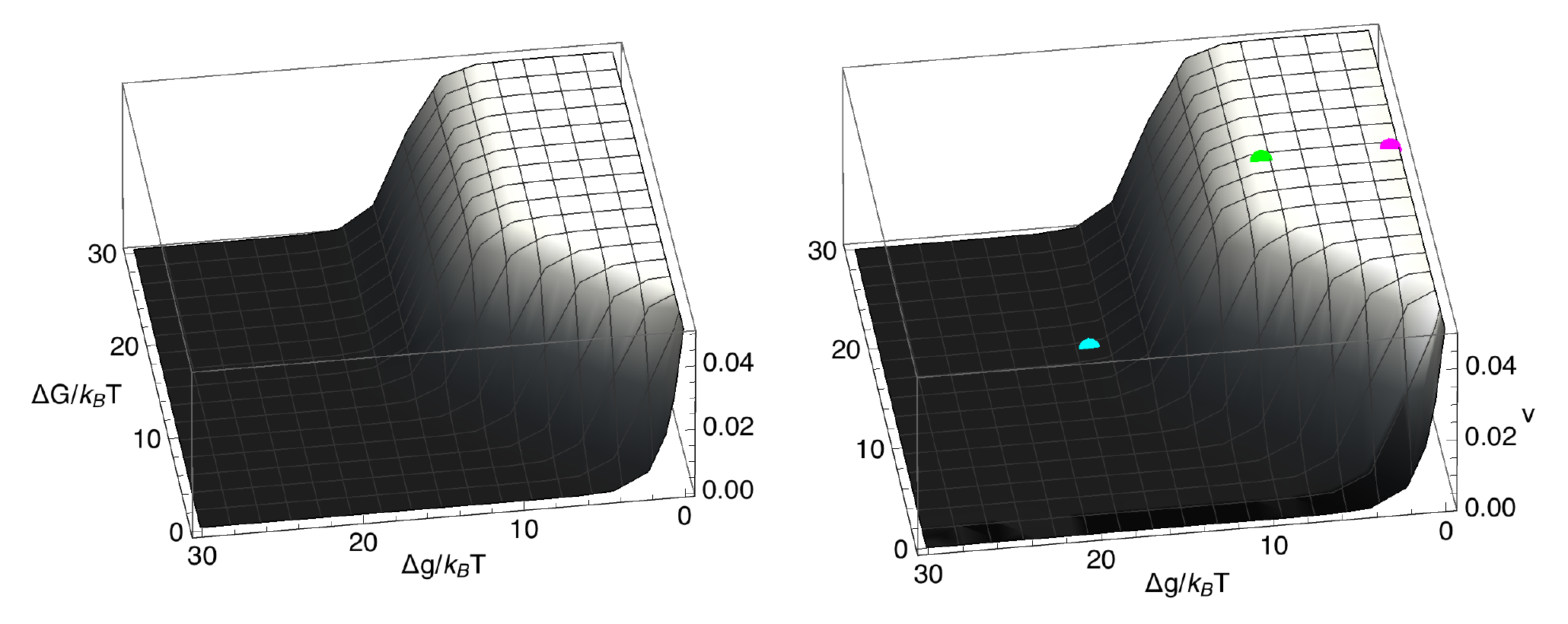}
\end{center}
\caption{
Mean velocity, $v$, of a composite LEF, plotted versus  $\Delta G/(k_BT)$ and $\Delta g/(k_BT)$
for
model 1 (left) and model 2 (right)
for $k_+ = 0.05$ per time step, $k_- = 5.0 \times 10^{-7}$ per time step,
$m_+ = 0.3$ per time step,   $m_- = 0.0003$ per time step, and (for model 2) $\alpha = 1$~per time step.
The cyan, green, and magenta points on the model-2 curve correspond to the theoretical mean velocities
of the composite LEFs whose positions versus time are shown in Fig.~\ref{Figure2}.
}
\label{Figure7}
\end{figure*}
In comparison to Equation~(\ref{EQ29}),
the velocity of a lonely remodeling complex, translocating on nucleosomal DNA,
unaccompanied by an SMC complex, is
\begin{eqnarray}
  v_R
&=  b k_+ P_1 -b k_- e^{-\frac{\Delta g}{k_B T}} (P_1+(1-P_1) e^\frac{\Delta G}{k_B T} ) \nonumber \\
&=b \frac{k_+-k_-}{1 +e^\frac{\Delta g-\Delta G}{k_BT}-e^{-\frac{\Delta G}{k_BT}}},
\label{Lonelyremodeling complex}
\end{eqnarray}
which may be straightforwardly obtained from Equation~(\ref{F18-VVV})
 by replacing $P_3$ with $e^{-\frac{\Delta g}{k_B T}}$,
which is the probability that there is a gap between the remodeler and junction 2.
The velocity of such a lonely remodeling complex
is relatively large for $\Delta G > \Delta g$ and is small otherwise.
Thus, as seems intuitive,  for efficient remodeler translocation on chromatin the remodeler-nucleosome repulsive free energy,
$\Delta G$
must exceed the free energy required for nucleosome unbinding, $\Delta g$.
In the large-$\Delta G$ limit,
the remodeler velocity realizes a plateau value of
\begin{equation}
v_R = b(k_+-k_-) = b k_-(\frac{k_+}{k_-} - 1),
\end{equation}
so that
the plateau velocity of a composite LEF
exceeds (is less than) [equals]
that of a lonely remodeling complex for $m_+ > m_-$ ($m_+ < m_-$) [$m_+ = m_-$].

We can also straightforwardly calculate the velocity of the SMC complex
 on nucleosomal DNA in the absence of the remodeling complex
 with the result that
 \begin{equation}
 v_S 
=  b e^{-\frac{\Delta g}{k_B T}} (m_+-m_-).
 \label{EQ18EQ18}
 \end{equation}
Equation~(\ref{EQ18EQ18}) informs us that, on nucleosomal DNA,
the velocity of loop extrusion by an isolated SMC complex,
which by assumption does not have its own nucleosome remodeling activity,
is suppressed by a factor $e^{-\frac{\Delta g}{k_B T}}$ compared to the velocity of its loop extrusion on nucleosome-free DNA,
which is $b(m_+-m_-)$.
Since $e^{-\frac{\Delta g}{k_B T}}$ is tiny, the velocity of the SMC without the remodeling complex
is correspondingly tiny, even for $m_+ \gg m_-$, emphasizing that the remodeling complex
is essential for significant loop extrusion in the chromatin context.
 
Equation~(\ref{EQ29}) informs us that the composite LEF's directionality
depends only on $\frac{k_+}{k_-}-\frac{m_-}{m_+}$.
Since we can expect  that $\frac{k_+}{k_-}=e^{-\frac{\Delta G_{R}}{k_BT}}$
and  $\frac{m_+}{m_-} = e^{-\frac{\Delta G_{S}}{k_BT}}$,
where $\Delta G_{R}$ is the free energy change associated with the remodeling complex stepping forward
and $\Delta G_{S}$ is the free energy change, associated with the SMC complex stepping forward,
it is clear that the composite LEF proceeds forward, only provided $\Delta G_{R}+\Delta G_{S} < 0$.
This outcome reflects the
Second Law of Thermodynamics,
 expressed
in the form that a chemical reaction proceeds forward only if the corresponding
change in free energy  is negative.
In comparison, Equation~(\ref{Lonelyremodeling complex}) informs us that a lonely remodeling complex
proceeds forwards if $\Delta G_R <0$.

Shown in Fig.~\ref{Figure7} is
a comparison between
 the LEF velocity for model 2
 and the LEF velocity for model 1.
Model 2 reproduces both the region in the $\Delta G$-$\Delta g$ plane where the composite LEF velocity is large
and the plateau value of the LEF velocity within that region [Equation~(\ref{plateau2})].
The cyan, green, and magenta points on the model-2 curve in Fig.~\ref{Figure7} correspond to the free energy settings and theoretical mean velocities
of the composite LEFs, whose simulated positions versus time are shown in Fig.~\ref{Figure2}.
Both the top group of traces and the middle group of traces in Figure~\ref{Figure2} fall within the plateau region of the velocity, which explains why their
velocities are very similar. However, while the middle LEF does fall within the plateau region of $P_3$, the top composite LEF exhibits a significantly
large value of $P_3$ and a correspondingly larger spatial extent.

The conceptually simplest versions of the composite LEF model (models 1 and 2) envision that
the remodeler ejects a nucleosome from the DNA ahead of the remodeler, and that the nucleosome subsequently rebinds behind the SMC.
Alternatively, model 3 hypothesizes an intermediate, ``remodeled'' state in which the displaced nucleosome  remains associated with the LEF,  eventually to relocate behind SMC.
This picture is reminiscent of the scenario envisioned in Ref.~\cite{Hodges:2009aa}, which demonstrated
experimentally that RNA polymerase
 could pass a nucleosome without causing nucleosome dissociation.
Nonetheless, 
for $e^\frac{\Delta g}{k_B T} \gg 1$, the predictions of all three models are indistinguishable.
The interpretation of $\Delta g$ is different for models 1 and 2, on the one hand,
and model 3 on the other. For models 1 and 2, $\Delta g$ is the nucleosome binding free energy,
which is several tens of $k_BT$.
For model 3, $\Delta g$ is the free energy of the remodeled configuration, relative to the free energy of a bound nucleosome, which we may expect to be smaller than the free energy required to nucleosome unbinding (models 1 and 2).
However, as noted above, the plateau value of the composite LEF's loop extrusion velocity is
independent of $\Delta g$ for all of the models.

\section{Conclusions}
\label{Conclusion}
A key result of this paper is that
even if nucleosomes block SMC translocation, efficient loop extrusion remains possible on chromatinized DNA
via a LEF, that
is a composite entity involving a remodeler and nucleosomes, as well as an SMC complex.
Thus, the possibility that nucleosomes may block SMC translocation and
loop extrusion on chromatin is {\em not} a reason to rule out the loop extrusion factor model of genome organization.

We have shown that, for a wide range of possible parameter values, such a
composite LEF exists and can give rise to loop extrusion with a velocity, that is comparable
to the remodeler's translocation velocity on chromatin, but is much larger than the velocity of a SMC complex
that is blocked by nucleosomes.
Although we have focused on one-sided loop extrusion,  two-sided loop extrusion simply requires two remodelers,
one for each chromatin strand threading the SMC. 

The composite LEF model is agnostic concerning whether the SMC complex shows ATP-dependent
translocase activity  ($m_+ \neq m_-$)
 or diffuses ($m_+ = m_-$) on naked DNA.
 However,  Equation~\ref{CompositeLEFCondition} specifies the condition for a composite LEF to exist
defined by the SMC and the remodeler being in close proximity,
 while
efficient chromatin loop extrusion  requires
repulsion between the  remodeler
and the junction between nucleosomal DNA and naked DNA,
that is large compared to the nucleosome binding free energy (models 1 and 2)
or the remodeled configuration free energy (model 3):
$\Delta G > \Delta g$.
An additional condition necessary for efficient loop extrusion is
$ m_+ > e^\frac{\Delta g}{k_BT} k_-$.
Finally, we remark that
the composite LEF model, described in this paper, is quite distinct from the models of 
Refs.~\cite{Brackley:2017aa,Maji2020},
which propose loop extrusion occurs without the involvement of a translocase.

\acknowledgements
This research was supported by NSF CMMI 1634988 and NSF EFRI CEE award EFMA-1830904. M. L. P. B. was supported by NIH T32EB019941 and the NSF GRFP.


%

\end{document}